\documentclass[letterpaper, 10 pt, conference]{ieeeconf}

\IEEEoverridecommandlockouts                              
\renewcommand{\baselinestretch}{1}

\usepackage[utf8]{inputenc}
\usepackage{amsmath}
\usepackage{amsfonts}
\usepackage{amssymb}
\usepackage{graphicx}
\usepackage{booktabs}
\usepackage{epstopdf}
\usepackage{mathrsfs}

\title{\LARGE \bf
Bayesian and regularization approaches to multivariable linear system identification: the role of rank penalties}
\author{G. Prando, A. Chiuso, G. Pillonetto
\\Dept. of Information Engineering, University of Padova, {\tt\small \{prandogi,chiuso,giapi\}@dei.unipd.it}
\thanks{This work has been partially supported by the FIRB project ``Learning meets time'' (RBFR12M3AC) funded by MIUR and by the European Community's Seventh Framework Programme [FP7/2007-2013] under
agreement n. 257462 HYCON2 Network of excellence.}
}

\begin{document}

\maketitle
\thispagestyle{empty}
\pagestyle{empty}

\begin{abstract}
Recent developments in linear system identification have proposed the use of non-parameteric methods, relying on regularization strategies, to handle the so-called bias/variance trade-off.  This paper introduces an impulse response estimator which relies on an  $\ell_2$-type regularization including a rank-penalty derived using the log-det heuristic  as a smooth  approximation to the rank function. This allows to account for different properties of the estimated impulse response (e.g. smoothness and stability) while also penalizing  high-complexity models. This also allows to account and enforce coupling between different input-output channels in MIMO systems. According to the Bayesian paradigm, the parameters defining the relative weight of the two regularization terms as well as the structure of the rank penalty are estimated optimizing the  marginal likelihood. Once these hyperameters have been estimated, the impulse response estimate is available in closed form.
Experiments  show that the proposed method is superior to the estimator relying on the ``classic'' $\ell_2$-regularization alone as well as those based in atomic and nuclear norm. 
\end{abstract}


\section{Introduction} \label{intro}

Linear system identification has been developed, by and large, following the so-called ``parametric approach'' \cite{Ljung:99,Soderstrom}. Candidate models within a prespecified model class (e.g. ARMAX/Box-Jenkins/State-space etc.) are parametrized using a finite dimensional parameter (say $\theta\in \Theta$). The most ``adequate'' parameter vector can be selected my minimizing a suitable cost function, most often the average squared prediction error resulting from the model (PEM). 
This approach heavily depends on the chosen model class and, in particular, on its complexity which will be called the \emph{system order} hereafter.

An alternative approach which has been put forward in the recent years \cite{Fazel01,SS2010,SS2011,ChenOL12,SurveyKBsysid,Recht2012AtomicnormCDC,SznaierECC2014} is based on the theory of regularization and Bayesian statistics;  factoring out minor differences, the regularization and Bayesian views just provide two alternative languages to describe the same approach.
 The basic idea is to choose a large enough (in principle infinite dimensional) model class so as to describe ``\emph{any}'' possible system and then introduce a penalty term (regularization view) or equivalently a prior probability (Bayesian view)
which is in charge of controlling the system complexity, thus facing the so called \emph{bias-variance} dilemma. 

In this paper we shall discuss, admittedly with a biased perspective due to our recent work, some Bayesian/regularization approaches which include  explicit penalties for system complexity derived from the Hankel matrix built with the impulse response coefficients.

The structure of the paper is as follows: Section \ref{PF} states the problem  and Section \ref{CM} describes the classic parametric approach. In Section \ref{sec:reg} we lie the basics of regularization which are then further developed in \ref{JR}. Algorithmic details are provided in Section \ref{sec:algorithm} while numerical results are provided in Section \ref{NE}.

\section{Problem Formulation}\label{PF}
We consider the following linear, causal and time-invariant (LTI) Output-Error (OE) system:
\begin{equation}\label{equ:sys}
y(t) = G(z) u(t) + e(t)
\end{equation}
where $y(t) = [y_1(t),..,y_p(t)]^\top \in\mathbb{R}^p$ is the $p$-dimensional output signal, $u(t)= [u_1(t),..,u_m(t)]^\top\in\mathbb{R}^m$ is the $m$-dimensional input signal, $e(t)$ is the innovation process and $G(z):={\cal Z}[g](z)$ 
is the system transfer function. For simplicity, we will assume the presence of a delay in $G(z)$, i.e. $G(\infty)=0$. In addition, we assume $e(t)\sim \mathcal{N}(0_p,\Sigma)$, $\Sigma=diag(\sigma_1,...,\sigma_p)$, with $\mathcal{N}(\mu,\sigma)$ being the Gaussian distribution with mean $\mu$ and variance $\sigma$; $0_p$ denotes the zero-vector of size $p$.

The objective is to estimate the impulse response coefficients $\left\{g(k)\right\}_{k\in\mathbb{Z}^+}$ from a finite set of input-output data $\left\{y(t),u(t)\right\}_{t\in[1,N]}$.

\section{Classical Methods}\label{CM}
Classical parametric approaches assume that $G(z)$ belongs to a certain parametric class and can be completely described by a parameter vector $\theta\in\mathbb{R}^n$, with $n$ denoting the model complexity; in this case we can adopt the notation $G_\theta(z)$. According to the prediction error methods (PEM), $\theta$ is estimated by minimizing the following function:
\begin{equation}\label{equ:j_p}
\widehat{\theta} = \arg \min_{\theta\in\mathbb{R}^n} J_P(\theta),\ \
J_P(\theta) = \sum_{t=1}^N \|y(t)-\hat{y}_{\theta}(t|t-1)\|^2
\end{equation}
where $\hat{y}_{\theta}(t|t-1)=G_\theta(z)u(t)$ denotes the one-step ahead predictor. An implicit assumption in the described procedure is that the number of available data $N$ is much larger than the model complexity $n$: in particular, many interesting properties have been derived for $N$ tending to infinity. However, PEM approaches are affected by some significant drawbacks. First, the optimization problem in \eqref{equ:j_p} becomes non-convex when certain model classes are chosen, giving rise to local minima issues. Second, the selection of the model complexity $n$ is a non-trivial step: for this purpose, many well-known tools exist, such as cross-validation or information criteria (AIC/FPE,BIC/MDL,etc.) \cite{Ljung:99,Soderstrom} but all of them require the estimation of many models with different complexities, thus significantly increasing the computational burden. Last, and most importantly, information criteria are derived from asymptotic arguments, whose validity is of course limited when dealing with a finite number $N$ of data. At last, the statistical properties of the obtained estimators $\widehat{\theta}$ are difficult to study \cite{Leeb} and experimental evidence \cite{SS2010,SS2011,ChenOL12} also shows unreliable results.

\section{Regularization approach}\label{sec:reg}
Recent developments in system identification have adopted regularization techniques, partly imported   from the machine learning and signal processing communities, in order to overcome the described issues. In particular, regularization allows to jointly perform estimation and model selection, by moving from families of rigid, finite parametric model classes to flexible, infinite dimensional models directly described by the impulse response $\left\{g_{\theta}(k)\right\}_{k\in\mathbb{Z}^+}$. In this case, the estimator $\widehat{\theta}$ is obtained as the solution of the following optimization problem
\begin{equation}\label{equ:reg_probl}
\widehat{\theta} = \arg \min_{\theta\in\mathbb{R}^n} J_P(\theta) + J_R(\theta)
\end{equation}
where $J_P(\theta)$ is the output data fit defined in \eqref{equ:j_p}, while $J_R(\theta)$ is a regularization term which penalizes certain parameters vectors $\theta$ which describe "unlikely" systems. Among the different forms of regularization $J_R(\theta)$ which have been proposed in the literature, two main classes can be identified: regularization for smoothness (aka Tikhonov regularization or Ridge regression) and regularization for selection. The first one gives rise to $\ell_2$-norm penalties, while the second one arises from convex relaxations of the $\ell_0$ quasi-norm (such as $\ell_1$ norm or its variations like the nuclear norm) or other non-convex sparsity inducing penalties. Inspired by the approach in \cite{Wipf12}, we propose an estimator which combines these two main types of regularization by exploiting both an $\ell_2$ norm penalty and a rank-penalty on the Hankel matrix of the estimated model. They are described in detail in Section \ref{sec:l2_reg} and \ref{sec:rank_reg}.

To simplify the derivation we consider a truncated impulse response $\left\{g_{\theta}(k)\right\}_{k\in\mathbb{Z}^+}$ of length $T$, $
G_{\theta}(z)=\sum_{k=1}^Tg_{\theta}(k)z^{-k}$,
where $T$ can always be taken large enough to catch the system dynamics.
Here, $\theta\in\mathbb{R}^{Tmp}$ is the parameter vector containing all the impulse response coefficients $\left\{g_{\theta}(k)\right\}_{k\in[1,T]}$, $g_{\theta}(k)\in\mathbb{R}^{p\times m}$, with the $ij$-th element $[g_{\theta}(k)]_{ij}$ being the $k$-th impulse response coefficient from input $j$ to output $i$:
\begin{eqnarray}
\theta &=& \left[\theta_{11}^\top\ \theta_{12}^\top \ \cdots \ \theta_{1m}^\top \ | \ \cdots \ | \ \theta_{p1}^\top \ \cdots \ \theta_{pm}^\top\right]^\top \label{equ:theta_ij}\\
\theta_{ij}&=&\left[[g_{\theta}(1)]_{ij}\ [g_{\theta}(2)]_{ij}\ \cdots \ [g_{\theta}(T)]_{ij}\right]^\top \nonumber
\end{eqnarray}
for $i= 1,...,p,\ j=1,...,m$. $^\top$ is the transpose operator.

We also introduce a vector notation by defining the vector of output observations, $Y\in \mathbb{R}^{Np}$,
\begin{equation}\label{equ:y}
Y = \left[y_1(1) \ \cdots \ y_1(N)\ | \ \cdots\ | \ y_p(1) \ \cdots \ y_p(N)\right]^\top
\end{equation}
and the regressors matrix $\Phi\in\mathbb{R}^{Np\times Tmp}$, $\Phi=\mbox{blockdiag}(\phi,\cdots,\phi)$, with
\vspace{-4mm}

\begin{footnotesize}
\begin{equation}\label{equ:phi}
\phi = \left[\begin{array}{ccc}
\varphi_1(1) &  \cdots & \varphi_m(1)\\
\vdots  & \ddots & \vdots \\
\varphi_1(N) & \cdots & \varphi_m(N)\\
\end{array}\right],\quad
\varphi_i(j) = \left[\begin{array}{c}
u_i(j-1) \\ u_i(j-2) \\ \vdots \\ u_i(j-T)
\end{array}\right]^\top
\end{equation}
\end{footnotesize}

\vspace{-3mm}

\noindent for $i =1,...,m,\ j=1,...,N$.
\\The cost function \eqref{equ:j_p} can now be formulated as
\begin{equation}
J_P(\theta) = \|Y-\widehat{Y}_{\theta}\|_2^2 = \|Y-\Phi\theta\|_2^2
\end{equation}
with $\widehat{Y}_{\theta}=\Phi\theta$ being the vectorized  one-step ahead predictor.

\subsection{Regularization for smoothness and stability}\label{sec:l2_reg}
This kind of regularization is derived in \cite{SS2010}, \cite{SS2011} by assuming that $\left\{g_{\theta}(k)\right\}_{k\in\mathbb{Z}^+}$ is a realization of a zero-mean Gaussian process with autocovariance $cov(g_{\theta}(i),g_{\theta}(j))=K(i,j)$. Since $K$ represents a Mercer Kerner, it is associated to a unique Reproducing Kernel Hilbert Space (RKHS) $\mathscr{H}$, to which $g_{\theta}$ is assumed to belong. When a finite impulse response $\left\{g_{\theta}(k)\right\}_{k\in[1,T]}$ is considered, the norm $\|g_{\theta}\|_{\mathscr{H}}$ defined in $\mathscr{H}$ can be expressed through a quadratic form:
\begin{equation}
\|g_{\theta}\|_{\mathscr{H}}^2 = \theta^\top K^{-1}\theta
\end{equation}
with $K\in\mathbb{R}^{Tmp\times Tmp}$, $[K]_{ij}=cov(g_{\theta}(i),g_{\theta}(j))$. Hence, the so-called $\ell_2$-type regularization is obtained by setting
$J_R(\theta) = \|g_{\theta}\|_{\mathscr{H}}^2 = \theta^\top K^{-1}\theta$.
The structure of $K$ can account for several properties, such as the fact that the impulse response of a linear system is an exponentially decaying function, or that it should be ``smooth''. See for instance \cite{SS2011,ChenOL12,SurveyKBsysid} for several choices, such as ``stable-spline'', diagonal, diagonal/correlated, tuned/correlated kernels. The specific structure of $K$ is defined through some hyperparameters $\alpha$ which can be estimated by cross-validation or by marginal-likelihood maximization (following the Empirical Bayes approach) \cite{SS2011}, \cite{Aravkin12}.

\subsection{Regularization for complexity}\label{sec:rank_reg}
Let us define the block Hankel matrix built with the impulse response coefficients of system \eqref{equ:sys}, $H(\theta) \in \mathbb{R}^{pr\times mc}$:
\begin{equation}\label{equ:hankel}
\textstyle
H(\theta)=\left[\begin{array}{cccc} g_{\theta}(1) & g_{\theta}(2) & \cdots & g_{\theta}(c)\\
g_{\theta}(2) & g_{\theta}(3) & \cdots & g_{\theta}(c+1)\\
\vdots & \vdots & \ddots & \vdots \\
g_{\theta}(r) & g_{\theta}(r+1) & \cdots & g_{\theta}(r+c-1)\end{array}\right]
\end{equation}
A classical result from realization theory  \cite{brockett1970finite} shows that the rank of the block Hankel matrix $H(\theta)$ equals the McMillan degree (i.e. the complexity $n$) of the system, if $r$ and $c$ are large enough. In this work $r$ and $c$ are chosen such that $r+c-1=T$ and the matrix $H(\theta)$ is as close as possible to a square matrix.

Hence, from the identification point of view, the complexity of the estimated model can be controlled by introducing a penalty on the rank of $H(\theta)$, i.e. by defining $J_R(\theta)=\mbox{rank}(H(\theta))$. However, this choice of $J_R(\theta)$ makes the optimization problem \eqref{equ:reg_probl} non-smooth and non-convex. To overcome this issue, in many previous works (\cite{Fazel01}, \cite{Chiuso13}) the rank penalty has been replaced by a penalty on its convex relaxation, i.e. the nuclear norm, by defining $J_R(\theta)=\|H(\theta)\|_*$. Recall that for a matrix $X$ the nuclear norm is defined by $\|X\|_*=\mbox{tr}(\sqrt{X^\top X})$. However, in this paper we adopt a non-convex approximation of the rank function, following what suggested in \cite{Wipf12} and \cite{Mohan10}. Recall that penalizing the rank of the Hankel matrix $H(\theta)$ is equivalent to favoring the sparsity of its singular values. A direct measure of sparsity in the components of a vector $x$ is given by its $\ell_0$ norm, $\|x\|_0$, which is equal to the number of non-zero components of $x$. Observing that
\begin{equation}
\sum_i \log|x_i| \equiv \lim_{p\rightarrow 0} \frac{1}{p}\sum_i(|x_i|^p-1) \propto \|x\|_0 
\end{equation}
we can approximate the $\ell_0$ norm of $x$ by its Gaussian entropy measure $\sum_i \log|x_i|$. Hence, in order to achieve sparsity in the singular values of $H(\theta)$, we define the penalty $J_R(\theta)=\log|H(\theta)H(\theta)^\top|=\sum_i\log \eta_i$, with $\left\{\eta_i\right\}_{i=[1,rp]}$ being the singular values of $H(\theta)H(\theta)^\top$. This type of penalty has been chosen since it can be reformulated in terms of a quadratic form in the vector $\theta$ (see Section \ref{sec:approx_log_det}): in a Bayesian setting this fact is exploited to define a Gaussian prior for $\theta$, as will be shown in Section \ref{sec:hyperp_est}.
 As in \cite{Hansson12} and  \cite{Chiuso13}, we also consider a weighted version of $H(\theta)$, i.e.
\begin{equation}
\widetilde{H}(\theta) = W_2^\top H(\theta)W_1^\top
\end{equation}
with $W_1$ and $W_2$ chosen so that the singular values of $\widetilde{H}(\theta)$ are conditional canonical correlation coefficients. Refer to \cite{Chiuso13} for a complete derivation of  $W_1$ and $W_2$. In particular, we adopted the second weighting scheme described in \cite{Chiuso13}.

\section{Regularization for both smoothness and complexity}\label{JR}
The estimator we propose is based on the combination of the two regularization types described in Sections \ref{sec:l2_reg} and \ref{sec:rank_reg}. We define
\begin{equation}\label{equ:mix_pen}
J_R(\theta) = \lambda_1\log|\widetilde{H}(\theta)\widetilde{H}(\theta)^\top|+\lambda_2\theta^\top K^{-1}\theta
\end{equation}
where $\lambda_1$ and $\lambda_2$ are non-negative scalar regularization parameters which control the relative weight of the two regularization terms. Details on how their value can be determined will be given in Section \ref{sec:hyperp_est}.

\textit{Remark:} Observe that in \eqref{equ:mix_pen} we have considered the weighted Hankel matrix $\widetilde{H}(\theta)=W_2^\top H(\theta)W_1^\top$. The following description will only refer to this more general case, since the non-weighted case can be easily recovered by setting $W_1=I_{cm}$ and $W_2=I_{rp}$. Here, $I_n$ denotes the identity matrix of size $n\times n$.

The impulse response coefficients contained in $\theta$ are then estimated by solving the following optimization problem:
\begin{eqnarray}
\widehat{\theta} = &\arg &\min_{\theta\in\mathbb{R}^{Tmp}}\left(Y-\Phi\theta\right)^\top\left(\Sigma^{-1} \otimes I_N\right)\left(Y-\Phi\theta\right)\nonumber\\
&+& \lambda_1\log|\widetilde{H}(\theta)\widetilde{H}(\theta)^\top|+\lambda_2\theta^\top K^{-1}\theta \label{equ:mix_reg_probl_1}
\end{eqnarray}
where $Y$ and $\Phi$ have been defined in \eqref{equ:y} and \eqref{equ:phi}, respectively. In \eqref{equ:mix_reg_probl_1} the available observations are also explicitly weighted by the inverse of the output noise variance, whose value has actually to be estimated, since it is not known a priori. In the simulations that follows its value has been set equal to the sample variance of the model obtained using only the $\ell_2$ type regularization (described in Section \ref{sec:l2_reg}).

Also, recall that the kernel $K$ depends on some hyper-parameters $\alpha$ which we consider fixed in this setting; for instance, they can be determined by cross-validation or by marginal likelihood maximization, as detailed in \cite{SS2010},\cite{SS2011} and \cite{ChenOL12}. The simulations we performed here exploit the latter procedure.

We now show how the optimization problem \eqref{equ:mix_reg_probl_1} can be solved by a sort of block-coordinate descent algorithm.

\subsection{Variational approximation to the log-det term}\label{sec:approx_log_det}
As a first step to formulate the block-coordinate descent algorithm, we need to determine a closed-form solution for the minimizer of the objective function in \eqref{equ:mix_reg_probl_1}. In this regard, observe that
the concave term $\log |\widetilde{H}(\theta)\widetilde{H}(\theta)^\top|$ can be expressed as the minimum of a set of upper-bounding lines \cite{Wipf12}:

\vspace{-5.5mm}

\begin{equation}\label{equ:logdet_bound}
\log |\widetilde{H}(\theta)\widetilde{H}(\theta)^\top| = \underset{\Psi\succ 0}{\min} \mbox{ tr}\left[\widetilde{H}(\theta)\widetilde{H}(\theta)^\top\Psi^{-1}\right]
+ \log |\Psi| - rp
\end{equation}
with $\Psi\in\mathbb{R}^{rp\times rp}$ being a positive definite matrix of so-called variational parameters. In addition, observe that the term $[\widetilde{H}(\theta)\widetilde{H}(\theta)^\top\Psi^{-1}]$ can be rewritten as a quadratic form in $\theta$. Indeed, letting $\Psi^{-1} = Q = LL^T$, we have
\begin{align}
\mbox{tr}\left[\widetilde{H}(\theta)\widetilde{H}(\theta)^\top Q\right]&= \mbox{tr}\left[L^\top \widetilde{H}(\theta)\widetilde{H}(\theta)^\top L\right]\label{equ:trace_hhq}\\
&= \|\mbox{vec}(\widetilde{H}(\theta)^\top L)\|_2^2\nonumber\\
&= \|(L^\top W_2^\top\otimes W_1)\mbox{vec}(H(\theta)^\top)\|_2^2\nonumber\\
&= \|(L^\top W_2^\top\otimes W_1)P\theta\|_2^2\nonumber\\
&= \theta^\top P^\top (W_2Q W_2^\top\otimes W_1^\top W_1)P\theta \label{equ:rank_quad_form}
\end{align}
where $P\in\mathbb{R}^{rpcm\times Tmp}$ is the matrix which vectorizes $H(\theta)^\top$, i.e. $\mbox{vec}\left(H(\theta)^\top\right) = P\theta$.
\\From \eqref{equ:logdet_bound} and \eqref{equ:rank_quad_form} we can upper bound the cost in \eqref{equ:mix_reg_probl_1} and re-define
\begin{align}
\widehat{\theta} =\arg \min_{\theta\in\mathbb{R}^{Tmp}}&\|\overline{Y}-\overline{\Phi}\theta\|_2^2 +\lambda_2\theta^\top K^{-1}\theta \nonumber\\
&+ \lambda_1\theta^\top P^\top (W_2Q W_2^\top\otimes W_1^\top W_1)P\theta  \label{equ:mix_reg_probl_3}
\end{align}
with $\overline{Y}=(\Sigma^{-1/2}\otimes I_N)Y$ and $\overline{\Phi}=(\Sigma^{-1/2}\otimes I_N)\Phi$.
\\For fixed $\lambda_1$, $\lambda_2$ and $Q$, the objective function in \eqref{equ:mix_reg_probl_3} is minimized in closed form by

\vspace{-3mm}

\begin{align}
\widehat{\theta} &= \left[\overline{\Phi}^\top\overline{\Phi}+A(Q,\lambda_1,\lambda_2)\right]^{-1}\overline{\Phi}^\top\overline{Y}
\label{equ:theta_min}\\
A(Q,\lambda_1,\lambda_2) &= \lambda_1P^\top (W_2Q W_2^\top\otimes W_1^\top W_1)P+\lambda_2K^{-1}\nonumber
\end{align}

\vspace{-1.5mm}

\noindent Next section will introduce a Bayesian perspective which allows to treat $\lambda_1$, $\lambda_2$ and $Q$ as hyper-parameters and to estimate them by marginal likelihood maximization.

\subsection{Hyper-parameters estimation}\label{sec:hyperp_est}
From a Bayesian point of view, the minimizer in \eqref{equ:theta_min} can be viewed as the MAP estimate of $\theta$ once defined the data distribution and prior:
\\$Y|\theta \sim \mathcal{N}\left(\Phi\theta,\Sigma\otimes I_N\right), \quad \scalebox{0.92}{$
\theta \sim \mathcal{N}\left(0_{Tmp},\left[A(Q,\lambda_1,\lambda_2)\right]^{-1}\right) $}
$\\

\vspace{0.5mm}

\noindent Observe that, exploiting the approximation to the log-det term described in \eqref{equ:logdet_bound} and to its reformulation as a quadratic form,  it is possible to define a Gaussian prior for $\theta$  as  in \eqref{equ:rank_quad_form}.

Within this Bayesian setting, the regularization coefficients $\lambda_1$, $\lambda_2$ and the matrix of variational parameters $Q$ can be treated as hyper-parameters; thus, following the Empirical Bayes Paradigm, they can be estimated by maximizing the marginal likelihood:
\begin{align}
\widehat{Q}, \widehat{\lambda_1}, \widehat{\lambda_2} &= {\arg\min}_{0\prec Q<1,\lambda_1,\lambda_2>0} \mathcal{L}\left(Q,\lambda_1,\lambda_2\right)\label{equ:ml_probl}\\
\mathcal{L}\left(Q,\lambda_1,\lambda_2\right) &= Y^\top \Lambda^{-1}Y + \log |\Lambda|
\end{align}
with $\Lambda = \Sigma\otimes I_N + \Phi \left[A(Q,\lambda_1,\lambda_2)\right]^{-1}\Phi^\top$.

Hence, once estimated $\widehat{Q},\ \widehat{\lambda}_1$ and $\widehat{\lambda}_2$ through \eqref{equ:ml_probl}, their values can be plugged in into \eqref{equ:theta_min} to find the desired estimate of $\theta$. Section \ref{sec:algorithm} will explain in detail how the estimation algorithm has been actually implemented.

\section{Algorithm implementation}\label{sec:algorithm}
As previously cited, the final estimate of $\theta$ is determined through a  block-coordinate descent algorithm which alternatively optimizes  $\widehat{\theta}$ using \eqref{equ:theta_min}  (which can be done in closed form for fixed $\widehat{Q},\ \widehat{\lambda}_1$, $\widehat{\lambda}_2$) and updates $\widehat{Q},\ \widehat{\lambda}_1$, $\widehat{\lambda}_2$  through \eqref{equ:ml_probl}. Our algorithmic implementation exploits the following variant of  \eqref{equ:ml_probl} to optimize  $\widehat{\lambda}_1$ and $\widehat{\lambda}_2$:
\begin{equation}\label{equ:ml_probl_fixedQ}
\widehat{\lambda}_1, \widehat{\lambda}_2 = {\arg\min}_{\lambda_1,\lambda_2>0} \mathcal{L}\left(\widehat Q,\lambda_1,\lambda_2\right)
\end{equation}
after $\widehat{Q}$ has been fixed based on  the current impulse response estimate (the details  will be given in Section \ref{sec:q_update}).

The iterations are stopped when the negative log likelihood  does not decrease. Note that no guarantees of achieving a local minimum can be given. 

Let $\widehat{\theta}^{(k)}$ denote the estimate of $\theta$ at the $k$-th iteration;   the notations $\widehat{Q}^{(k)},\ \widehat{\lambda}_1^{(k)},\ \widehat{\lambda}_2^{(k)}$ will have analogue meaning. The algorithm can be summarized as follows:
\begin{enumerate}
\item Set $K$ to be the kernel estimated using marginal likelihood optimization when no rank penalty is included, as done in \cite{SS2010}.
\item Set $\widehat{\theta}^{(0)}$ equal to the estimate obtained using only the $\ell_2$ type regularization.
\item Define $\widehat{Q}^{(0)}$ using the procedure  in Section \ref{sec:q_update}.
\item Determine $\widehat{\lambda}_1^{(0)}$ and $\widehat{\lambda}_2^{(0)}$ solving \eqref{equ:ml_probl_fixedQ} with $\widehat Q=\widehat{Q}^{(0)}$ .
\item At the $(k+1)$-th iteration
\begin{itemize}
\item Compute $\scriptstyle\widehat{\theta}^{(k+1)}=\left[\overline{\Phi}^\top\overline{\Phi}+A\left(\widehat{Q}^{(k)},\widehat{\lambda}_1^{(k)}, \widehat{\lambda}_2^{(k)}\right)\right]^{-1}\overline{\Phi}^\top\overline{Y}
$
\item Determine $\widehat{Q}^{(k+1)}$ as in Section \ref{sec:q_update}.
\item Update $\widehat{\lambda}_1^{(k+1)}$ and $\widehat{\lambda}_2^{(k+1)}$ solving \eqref{equ:ml_probl_fixedQ} with $\widehat Q=\widehat{Q}^{(k+1)}$.
\item Stop to iterate if $\mathcal{L}(\widehat{Q}^{(k+1)},\widehat{\lambda}_1^{(k+1)},\widehat{\lambda}_2^{(k+1)}) \geq \mathcal{L}(\widehat{Q}^{(k)},\widehat{\lambda}_1^{(k)},\widehat{\lambda}_2^{(k)})$ and choose $\widehat{\theta}^{(k)}$ as the final $\theta$ estimate.
\end{itemize}
\end{enumerate}

\textit{Remark}:  Experimental evidence shows that, when optimizing w.r.t. $\lambda_1$ and $\lambda_2$, it is convenient to set a lower bound on the value of $\lambda_2$.
As a result, for instance, stability of the estimator is preserved. The constraints on the hyperparameters can be seen as an hyper regularizer which limits the degrees of freedom due to hyperparameter estimation \cite{PCECC2014}. 
 
\subsection{Update of the matrix $Q$}\label{sec:q_update}
Let us consider equation \eqref{equ:trace_hhq} and let $\widetilde{H}(\theta)=U(\theta)S(\theta)V(\theta)^\top$ denote the singular value decomposition of $\widetilde{H}(\theta)$, with $S(\theta)=diag(s_1,...,s_{pr})$. To simplify the notation, in the following we will omit the dependence of $U$, $S$ and $V$ on $\theta$. We can rewrite \eqref{equ:trace_hhq} as follows:

\vspace{-2mm}

\begin{equation}\label{equ:trace_hhq_2}
\mbox{tr}\left[\widetilde{H}(\theta)\widetilde{H}(\theta)^\top Q\right] = \mbox{tr}\left[US^2U^\top Q\right]
\end{equation}

\vspace{-1mm}

\noindent Let $Q=U_QS_QV_Q^\top$ be the singular value decomposition of $Q$, with $S_Q=diag(s_1^Q,...,s_{pr}^Q)$; if we set $U_Q=V_Q=U$, from \eqref{equ:trace_hhq_2} we have
$\mbox{tr}[\widetilde{H}(\theta)\widetilde{H}(\theta)^\top Q] = \mbox{tr}[S^2S_Q] $.
Recalling that the term $\mbox{tr}[\widetilde{H}(\theta)\widetilde{H}(\theta)^\top Q]$ is included in the regularization function $J_R(\theta)$, from this last equation we can see that the singular values of $Q$ act as penalties on the squares of the singular values of $\widetilde{H}(\theta)$: the larger the first ones, the smaller will be the estimated latter ones. 

If the Hankel matrix $\widetilde{H}(\theta_0)$ of the true system was known, a natural choice for $Q$ would be $U_Q=U_0$ and $s_i^Q=1/s_{0,i}^2$, where $s_{0,i}$ denotes the $i$-th singular value of $\widetilde{H}(\theta_0)$. However, since at each iteration of the algorithm described in Section \ref{sec:algorithm} an impulse response estimate $\widehat{\theta}^{(k)}$  is available, we can exploit it to build $\widehat{Q}^{(k)}$. Namely, letting  $\widetilde{H}(\widehat{\theta}^{(k)})=U^{(k)}S^{(k)}V^{(k)^\top}$  with $S^{(k)}=diag(s_1^{(k)},...,s_{rp}^{(k)})$ being  the singular value decomposition of $\widetilde{H}(\widehat{\theta}^{(k)})$, we set:

\vspace{-2mm}

\begin{equation}\label{equ:q_hat}
\widehat{Q}^{(k)} = U^{(k)}\ diag\left(1/(s_1^{(k)})^2,...,1/(s_{rp}^{(k)})^2\right)U^{(k)^\top}
\end{equation}

\vspace{-1mm}

\noindent By means of the simulations we performed, we observed that the singular values of $\widehat{Q}^{(k)}$ as defined  in \eqref{equ:q_hat} may become very large, probably providing excessive prior along certain directions (the columns of $U^{(k)}$ which, we recall, is  estimated and thus subject to uncertainty).
Thus we found that a less committing prior, which just gives the same weight to all small singular values below a certain threshold, is to be preferred. 
In order to identify this threshold we recall that the matrix $\widetilde{H}(\widehat{\theta}^{(k)})$  has been formed using sample covariances (see Section  \ref{sec:rank_reg}). Thus we recall a result on the uniform convergence of sample covariances (see e.g. Theorem  5.3.2 in \cite{HannanDeistler}), which shows that, under mild assumptions, the difference between the true and the 
estimated sample covariance is (uniformly in the lag) of the order of $O\left(\sqrt{\frac{\log(\log(N))}{N}}\right)$, where $N$ is the number of available observations.  Therefore, the first ``noise'' singular value of $ \widetilde{H}(\widehat{\theta}^{(k)}) \widetilde{H}(\widehat{\theta}^{(k)})^\top$ (i.e. that corresponding to the noise subspace,   orthogonal to the image of 
$\widetilde {H}({\theta_0})\widetilde {H}({\theta_0})^\top$), is expected to be of the size $O\left(  \frac{\log(\log(N))}{N}\right)$.
  Thus it is to be expected that singular values above that threshold are due to ``signal'' components while the smaller ones may be corrupted by noise. 
Hence, we re-define the singular values $s_i^{\widehat{Q}^{(k)}}$ of $\widehat{Q}^{(k)}$ as
\begin{equation}
s_i^{\widehat{Q}^{(k)}} = \left\{ \begin{array}{ll} \left(s_i^{(k)}\right)^{-2} \quad &\mbox{if } s_i^{(k)}\geq \sqrt{c \frac{\log(\log(N))}{N}}\\
\nu(N) &\mbox{if } s_i^{(k)}<  \sqrt{c {\frac{\log(\log(N))}{ N}} }\end{array} \right.
\end{equation} 
where  $c$ is  a constant which, in the simulation results, we have taken equal to the number of rows of the Hankel matrix.  The saturation value $\nu(N)$ is defined as
 $\nu(N) := {\frac{10N}{c \log(\log(N))}}$. Thus, we replace the update in \eqref{equ:q_hat} by
\begin{equation}
\widehat{Q}^{(k)} = U^{(k)}\ diag\left(s_1^{\widehat{Q}^{(k)}},...,s_{rp}^{\widehat{Q}^{(k)}}\right)U^{(k)^\top}
\end{equation}

\section{Numerical experiments}\label{NE}
We now test and compare the proposed estimator on some Monte Carlo studies, generated under three scenarios, S1, S2 and S3, described below. In all cases the innovation process $e(t)$ is a zero-mean white noise with standard deviation chosen randomly at each run in order to guarantee that the signal to noise ratio on each output channel is a uniform random variable in the interval $[1,4]$ for S1 and S2 and $[1,10]$ for S3.

\begin{itemize}
\item[{\bf S1})] We consider a fixed fourth order system with transfer function $G(z)  = C(zI-A)^{-1}B$ where 
$$
\begin{array}{c}
A = \mbox{blockdiag}\left(\left[\begin{array}{cccc}.8 &.5 \\ -.5  & .8\end{array} \right],\left[\begin{array}{cccc}.2 &.9 \\ -.9  & .2\end{array} \right]\right) \\
B = \left[ 1\; 0\; 2 \; 0\right]^\top  \quad C = \left[\begin{array}{cccc}  1 & 1 & 1 & 1 \\ 0 & .1 & 0 & .1 \\ 20 & 0 & 2.5 & 0\end{array} \right]
\end{array}
$$
The input is generated, for each Monte Carlo run, as a low
pass filtered white noise with normalized band $[0,\zeta]$ where  $\zeta$ is a uniform random variable in the interval $[0.8, 1]$. $N_{MC1}=200$ Monte Carlo runs are considered.
\item[{\bf S2})] For each Monte Carlo run  $G(z)$ is generated randomly using the Matlab function {\tt drmodel} with $3$ outputs and $1$ input   while guaranteeing that all the poles of $G(z)$ are inside the disc of radius $.85$ of the complex plane. System orders are randomly chosen from 1 to 10. The input $u(t)$ is zero-mean unit variance white noise (similar behavior is obtained with low pass noise, not reported here for reasons of space). $N_{MC2}=200$ Monte Carlo runs are considered.
\item[{\bf S3})] For each Monte Carlo run a SISO continuous-time system is generated using the Matlab function \verb!rss!. System order is randomly chosen from 1 to 30. Each continuous-time system is sampled at 3 times the bandwith in order to derive the corresponding discrete-time system. The input $u(t)$ is zero-mean unit variance white noise filtered through a randomly generated second order filter. $N_{MC3}=120$ Monte Carlo runs are considered.
\end{itemize}
We now consider the following algorithms:
\begin{enumerate}
\item ATOM: The estimator proposed in \cite{Recht2012AtomicnormCDC} which adopts a regularization based on the atomic norm of the transfer function to be estimated. The algorithm run on a set of $h \times k = 32 \times 29 = 928$ atoms built from the impulse responses of second order linear systems 
$$
G_{hk}(z) = C\frac{z}{(z-p_{hk})(z-p^*_{hk})} \quad \quad p_{hk} = \rho_h e^{j\theta_k}
$$

\vspace{-4mm}

\begin{align}
\mbox{where} \quad \rho_h &\in [0.41:0.02:0.99 \; 0.995\; 0.999]  \in \mathbb{R}^{32}\nonumber\\
\theta_k &\in [(\pi/30):(\pi/30):(\pi-\pi/30)] \in \mathbb{R}^{29} \nonumber
\end{align}
and $C$ is determined to guarantee unit norm. We rely on the  \verb!glmnet!  package \cite{glmnet} for the numerical implementation. This algorithm is tested only on the SISO scenario S3.
\item PEM: The classical PEM approach, as implemented in the \verb!pem.m! function of the MATLAB System Identification toolbox.
\item SS: The stable-spline estimator developed in \cite{SS2010} and \cite{ChenOL12}, applied independently on each output channel. First order stable splines are used in scenario S1 and S3, while second order stable splines are adopted in S2. For both scenarios an ``OE'' model class is assumed.
\item SSNN: The estimator proposed in \cite{Chiuso13} which combines the $\ell_2$-type penalty of Section \ref{sec:l2_reg} (with kernel estimated by the SS algorithm in 2) with a nuclear norm penalty on the weighted Hankel matrix $\widetilde{H}(\theta)$. The weights $W_1$ and $W_2$ are computed as illustrated in \cite{Chiuso13}. The regularization parameters $\lambda_1$ and $\lambda_2$ are estimated through cross-validation on a  predefined grid with $10$ values for each hyperparameter; the ``location'' of the grid is chosen to optimize performance.
\item SSR: The estimator \eqref{equ:mix_reg_probl_1} obtained through the algorithm described in Section \ref{sec:algorithm}. Both the Hankel matrix $H(\theta)$ in \eqref{equ:hankel} and its weighted version $\widetilde{H}(\theta)$ are considered and compared.
\end{enumerate}
The complexity of the models estimated through PEM is the default one provided by MATLAB, while for the other algorithms the length $T$ of the estimated impulse response was set to 80 for S1, 50 for S2 and 60 for S3.
All the considered estimators are obtained using $N=500$ pairs of input-output data for scenarios S1 and S2, while $N=1000$ data are used for S3. Their performances are compared by evaluating the \textit{Average Impulse Response Fit}, such defined:
\begin{equation}\label{equ:fit_imp}
\resizebox{\columnwidth}{!}{$
\mathcal{F}(\widehat{\theta}) = \frac{1}{pm}\sum_{i,j} 100 \left(1-\frac{\|\theta_{ij}^0-\widehat{\theta}_{ij}\|}{\|\theta_{ij}^0 - \bar{\theta}_{ij}^0\|} \right), \ \bar{\theta}_{ij}^0 = \frac{1}{T}\sum_{k=1}^T g^0_{ij}(k)
$}
\end{equation}
where $\theta_{ij}$ has been defined in \eqref{equ:theta_ij}, while $\theta_{ij}^0$ contains the true coefficients $\{g_{ij}^0(k)\}_{k=1,..,T}$ of the impulse response from input $j$ to output $i$.


\subsection{Results}
\begin{figure}
\centering
\includegraphics[width=\columnwidth]{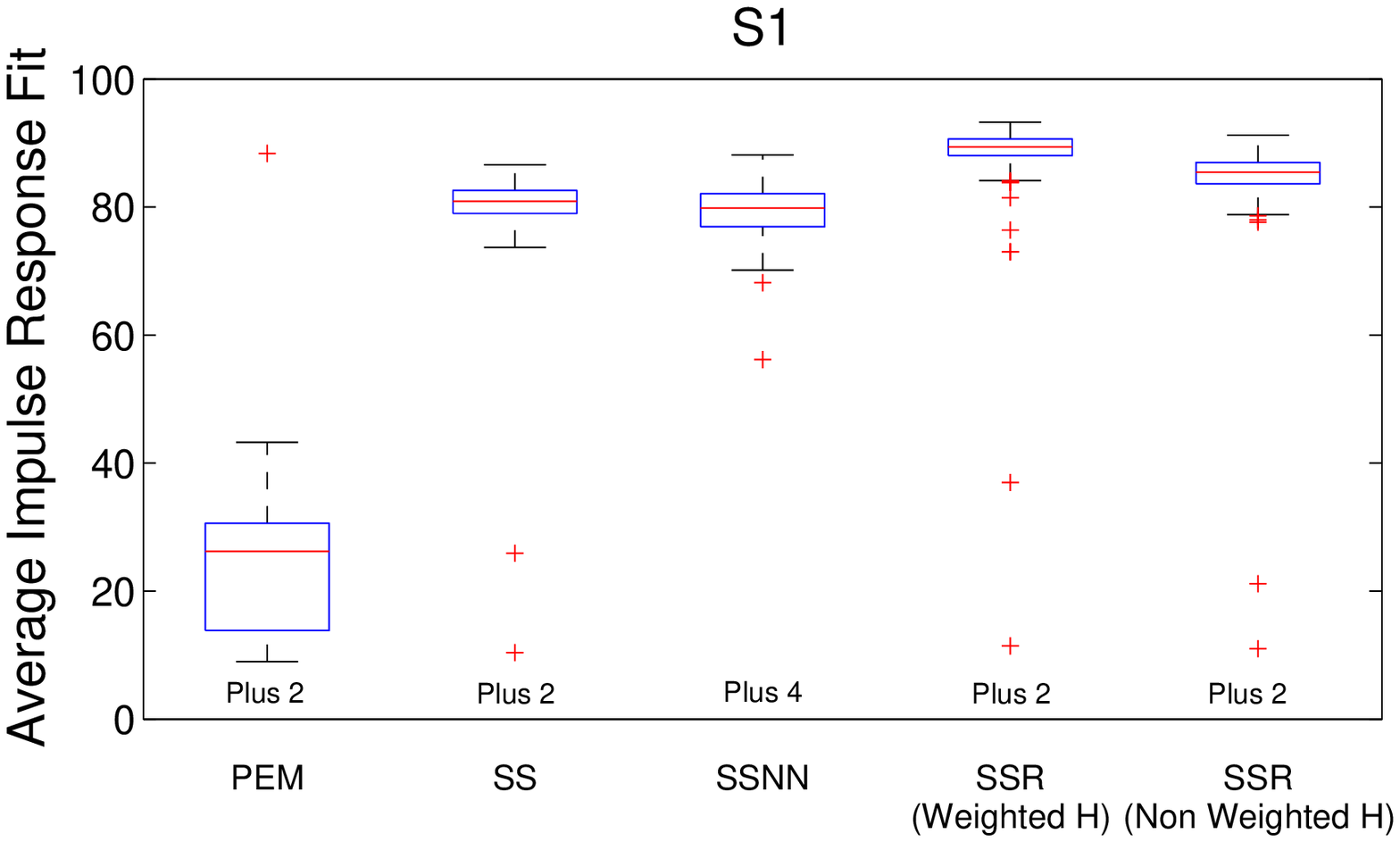}
\includegraphics[width=\columnwidth]{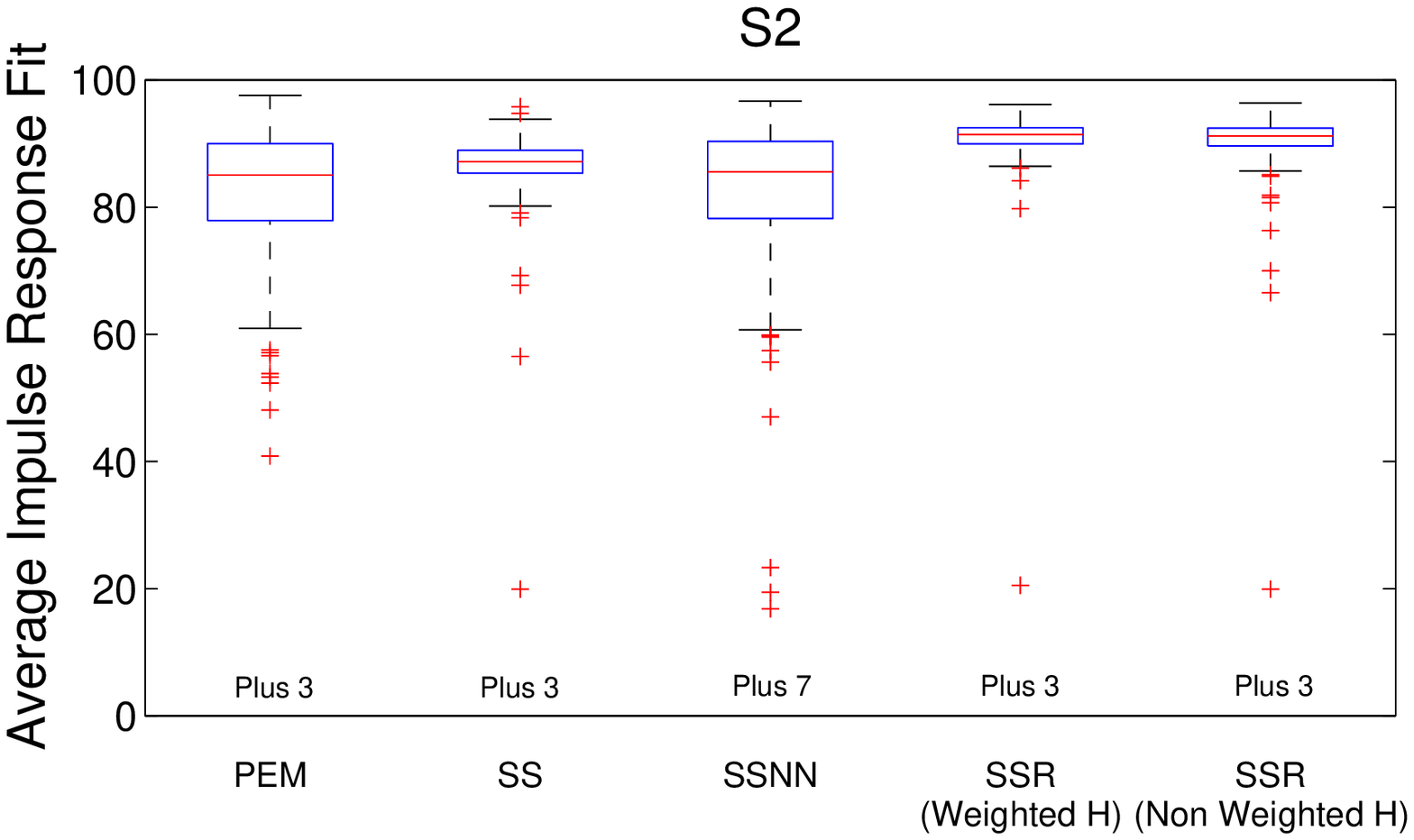}
\includegraphics[width=\columnwidth]{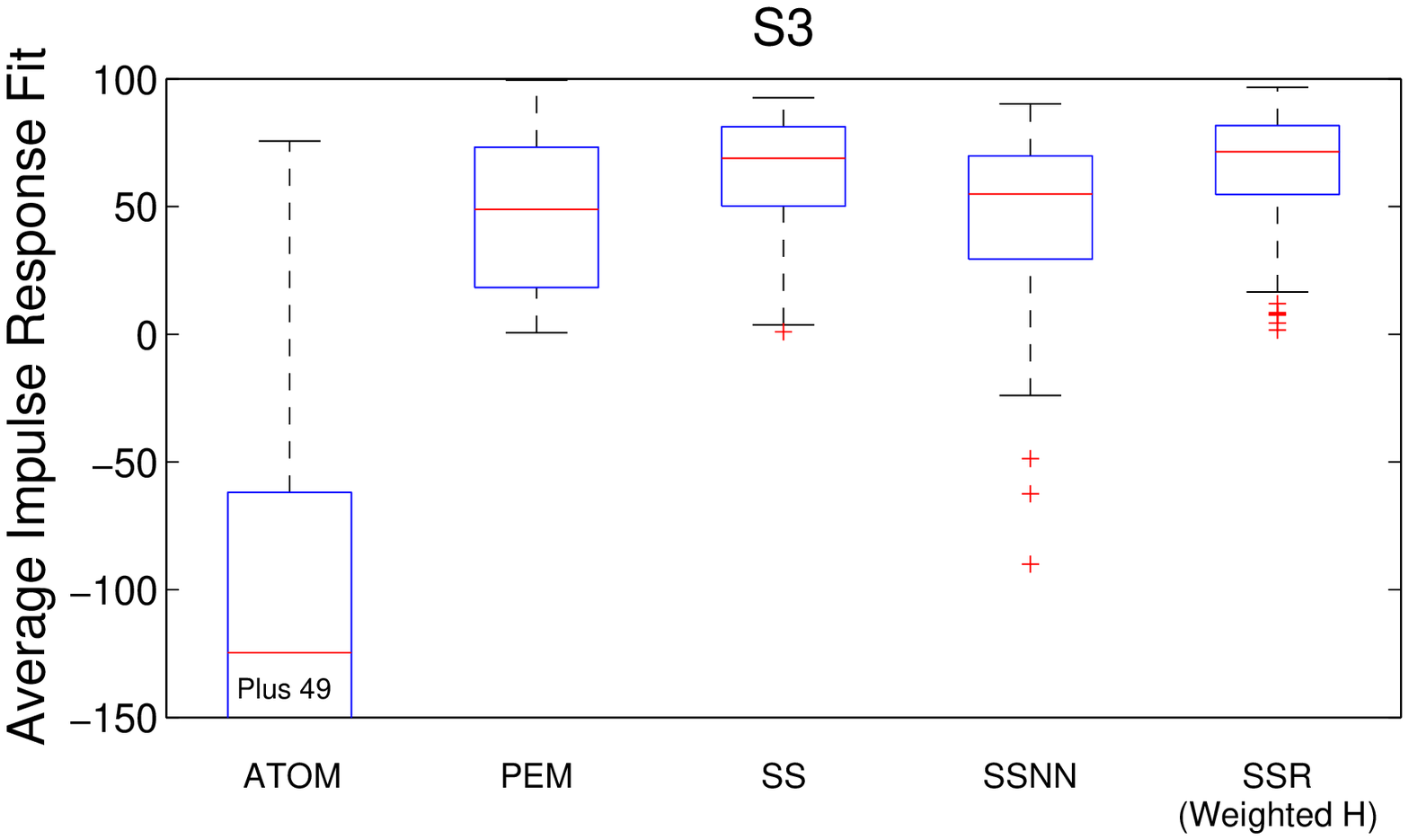}
\vspace{-7mm}
\caption{Boxplot of the Average Impulse Response Fit for the three experimental scenarios S1, S2 and S3.}\label{fig:fit_imp}
\vspace{-2mm}
\end{figure}

The boxplots in Figure \ref{fig:fit_imp} prove the effectiveness of the proposed estimator: in the three experimental setups here considered, it outperforms the other pre-existing methods, even if SS gives  comparable performances on S3. In particular, its implementation with the weighted version of the Hankel matrix, $\widetilde{H}(\theta)$, seems preferable to the non-weighted one. 

The tests performed on S3 also show how the regularization based on the atomic norm may lead to unreliable results, compared to the other approaches here evaluated.


\section{Conclusion and Future Work}
We have presented a novel regularization-based approach for linear system identification. The $\ell_2$ penalty  combines two types of regularization, thus jointly enforcing "smoothness" through a ``stable-spline'' penalty as well as``low-complexity" through a relaxation of a rank penalty on the Hankel matrix. The reformulation of the rank penalty as an $\ell_2$ cost  provides a Bayesian interpretation of the proposed estimator, which, in turn, can be used to perform hyperparameter selection via Marginal Likelihood maximization. 

Simulation results show  the performance improvements achievable through this novel method w.r.t. to other pre-existing approaches.

In our future work we plan to further investigate the properties of the kernel $A(Q,\lambda_1,\lambda_2)^{-1}$ which characterizes the new $\ell_2$ penalty  derived in this paper as well as to study the relation with other identification algorithms based on complexity penalties. In addition, different ways to update the matrix $Q$ will be explored and compared to the one here proposed.

\begin{table}
\caption{Medians of $\mathcal{F}(\widehat{\theta})$ in scenarios S1, S2 and S3.}
\vspace{-4mm}
\centering
\begin{tabular}{cccccc}
\toprule
& PEM & SS & SSNN & SSR ($\widetilde{H}(\theta)$) & SSR ($H(\theta)$) \\
\midrule
S1 & 24.88 & 80.90 & 79.89 & 89.34 & 85.46 \\
S2 & 83.40 & 87.15 & 80.20 & 91.40 & 91.20 \\
S3 & 48.94 & 68.93 & 54.88 & 71.53 & - \\ 
\bottomrule
\end{tabular}
\end{table}

\sectionmark{References}
\nocite{*}
\bibliographystyle{IEEEtran}
\bibliography{IEEEabrv,References}

\end{document}